\documentclass[10pt]{wlscirep}
\usepackage[T1]{fontenc}
\usepackage[utf8]{inputenc}
\usepackage[caption=false]{subfig}
\usepackage{amsmath}
\usepackage{hyperref}
\usepackage[sort&compress,numbers]{natbib}
 
\title{Quasi-quantized Hall response in bulk InAs}

\author[1,*]{R. Wawrzy\'{n}czak}
\author[1]{S. Galeski}
\author[1]{J. Noky}
\author[1]{Y. Sun}
\author[1]{C. Felser}
\author[1,2,**]{J. Gooth}
\affil[1]{Max Planck Institute for Chemical Physics of Solids, 01187 Dresden, Germany}
\affil[2]{Institut für Festkörper- und Materialphysik, Technische Universität Dresden, 01062 Dresden, Germany}

\affil[*]{rafal.wawrzynczak@cpfs.mpg.de}
\affil[**]{johannes.gooth@cpfs.mpg.de}

\begin{abstract}
The quasi-quantized Hall effect~(QQHE) is the three-dimensional~(3D) counterpart of the integer quantum Hall effect~(QHE), exhibited only by two-dimensional~(2D) electron systems. It has recently been observed in layered materials, consisting of stacks of weakly coupled 2D platelets that are yet characterized by a 3D anisotropic Fermi surface. However, it is predicted that the quasi-quantized 3D version of the 2D QHE should occur in a much broader class of bulk materials, regardless of the underlying crystal structure. Here, we compare the observation of quasi-quantized plateau-like features in the Hall conductivity of the $n$-type bulk semiconductor InAs with the predictions for the 3D QQHE in presence of parabolic electron bands. InAs takes form of a cubic crystal without any low-dimensional substructure. The onset of the plateau-like feature in the Hall conductivity scales with $\sqrt{2/3}k_{\textrm{F}}^{z}/\pi$ in units of the conductance quantum and is accompanied by a Shubnikov-de Haas minimum in the longitudinal resistivity, consistent wit the results of calculations. This confirms the suggestion that the 3D QQHE may be a generic effect directly observable in materials with small Fermi surfaces, placed in sufficiently strong magnetic fields.
\end{abstract}
\begin{document}

\flushbottom
\maketitle

\thispagestyle{empty}

\section*{Introduction}

Electrons subjected to a magnetic field~($B$), are forced to move on curved orbits with a discrete set of energy eigenvalues - the Landau levels~(LLs). By increasing $B$, cyclotron orbits are getting narrower and the LLs move upwards in energy and cross the Fermi level $E_{\textrm{F}}$ one after another. Depopulation of energy levels pushed above the electrons' chemical potential results in abrupt changes in charge carriers' density of states in the vicinity of $E_{\textrm{F}}$, reflected in oscillatory behavior of transport and thermodynamic quantities~(quantum oscillations)~\cite{shoenberg1984}. At sufficiently large $B$, where only a few LLs are occupied, 2D electron systems enter the quantum Hall regime. It is characterized by a fully gapped electronic spectrum in the bulk and current-carrying gapless edge states, in which the Hall conductance $G_{xy}$ becomes precisely quantized in units of the inverse of the von Klitzing constant, \textit{i.e.} half of the conductance quantum $1/R_{\textrm{K}}=G_{0}/2=e^2$/$h$, value based solely on the fundamental constants: the electron charge $e$ and the Planck constant $h$~\cite{klitzing1980}. This is the quantum Hall effect, traditionally considered to be strictly limited to 2D electron systems.

Following the nature of QHE, plateau-like features in Hall response, observed in various systems, are eagerly ascribed to non-trivial topology of the band structure. However, in 3D systems, the exact quantization of the Hall conductance is deterred due to the dispersion of the LL bands in the third dimension, preventing the opening of the bulk gap, which is an important ingredient that leads to quantized values of Hall conductivity~($\sigma_{xy}$). Efforts to extend the QHE to 3D systems usually involve an additional mechanism, that transforms the original 3D system into spatially separated 2D quantum Hall layers, stacked along the magnetic field direction. This reduces the problem to the parallel conduction of decoupled 2D electron systems, each of them being in the quantum Hall regime. The total Hall conductance of the 3D system is then given by discrete values, the multiples of $e^{2}/h\cdot{}N$, where $N$ is the numbers of created 2D layers. Material systems to realize this version of the 3D QHE are metals, semimetals and doped semiconductors, in which a periodic potential modulation is imposed by the lattice structure~\cite{bernevig2007,stormer1986}; by spin and charge density waves~\cite{halperin1987,mckernan1995,balicas1995}; or by standing electron waves in sufficiently thin samples~\cite{yin2019}. Interestingly, in several Weyl and Dirac-semi metals a quasi-quantized Hall effect was observed, where quantum oscillations in the Hall voltage closely mimic the Hall response of 2DEG systems~\cite{shekhar2015,Tang2019,galeski2020,galeski20202}. In these cases, the Hall conductivity does not take precisely quantized values, but it involves a consistent, well-defined scaling prefactor $k_{0}$ and is called \textit{quasi-quantized}.

Indeed, plateau-like features have been observed in the Hall response at the lowest Landau level of the 3D Dirac semimetals ZrTe$_{5}$~\cite{Tang2019,galeski20202} and HfTe$_{5}$~\cite{galeski2020}. In these materials, the Hall conductivity $\sigma_{xy}$ scales with $e^{2}/h\cdot{}k_{0}/\pi{}$, where for magnetic field applied along $z$, $k_{0}=k_{\textrm{F}}^{z}$, with $k_{\textrm{F}}^{z}$ being the Fermi wave vector component in the direction of $B$~\cite{Tang2019,galeski20202}. However, ZrTe$_{5}$ and HfTe$_{5}$ are in fact layered materials with a highly anisotropic Fermi surface (aspect ratio 1:10) and are, hence, on the verge of a 2D system. The initial experimental observations were concluded to be a result of an instability of Fermi surface, i.e. charge density wave~(CDW)~\cite{Tang2019}, establishing a stack 2D QHE layers and thereby lowering the dimensionality of electronic subsystem. This interpretation assigned the observed scaling of conductance being related to the characteristic wave-vector of the CDW distortion. However, further investigation have shown lack of any signatures expected in the presence of CDW transition and the field-dependence of the thermodynamic and transport properties might be extracted from the calculated linear response of the 3D band model~\cite{galeski20202}. The calculations have shown that, due to low charge carrier concentration the particle number tends to stay fixed forcing the, more energetically favorable, variations of electrons' chemical potential responsible for the observed anomalies. 

 Furthermore, theoretical study applying similar approach to free electron model finds $k_{0}$ given by multiples of $\sqrt{\frac{2}{3}}k_{\textrm{F}}^{z}$~\cite{noky2020}. In both cases the characteristic length scale is strictly related to the Fermi wavelength ${\lambda}_{\textrm{F}}^{z}$ = 2$\pi$/$k_{\textrm{F}}^{z}$ along the field direction, marking a possibility of strong resemblance, between both observations. Experimentally, the $k_{0}$-scaling in the vicinity of the quantum limit~(QL) of 3D materials has so far been shown only for Dirac systems. It is therefore desirable to go beyond these experiments and investigate the Hall effect below the QL of isotropic 3D electron systems with parabolic bands where, after appropriate readjustments, the results for free electron might be still applicable. 

In this regard, many of III–V semiconductor materials are particularly interesting (\textit{e.g.} InAs, InSb, GaAs), as they are known for having a cubic lattice structure, high charge carrier mobilities and almost parabolic bands. What is most important is that their Fermi level, with use of chemical doping, or by spontaneous defects, can be placed just above the edge of the conduction band allowing to reach the QL in moderate magnetic fields. In fact, previous measurements of the Hall resistivity on InAs and InSb bulk crystals revealed oscillatory features of the Hall coefficient in the vicinity of the lowest Landau level~\cite{pavlov1965,mani1990,zeitler1994}. In order to investigate this in the light of the recent developments, we have chosen undoped $n$-type InAs as a material, which is accessible with extremely low charge carrier concentrations and relatively high electron mobilities.

\section*{Results}

\begin{figure}[ht]
\subfloat{\label{fig:fig1a}}
\subfloat{\label{fig:fig1b}}
\subfloat{\label{fig:fig1c}}
\subfloat{\label{fig:fig1d}}
\subfloat{\label{fig:fig1e}}
\subfloat{\label{fig:fig1f}}
\subfloat{\label{fig:fig1g}}
\subfloat{\label{fig:fig1h}}
\includegraphics[width=\linewidth]{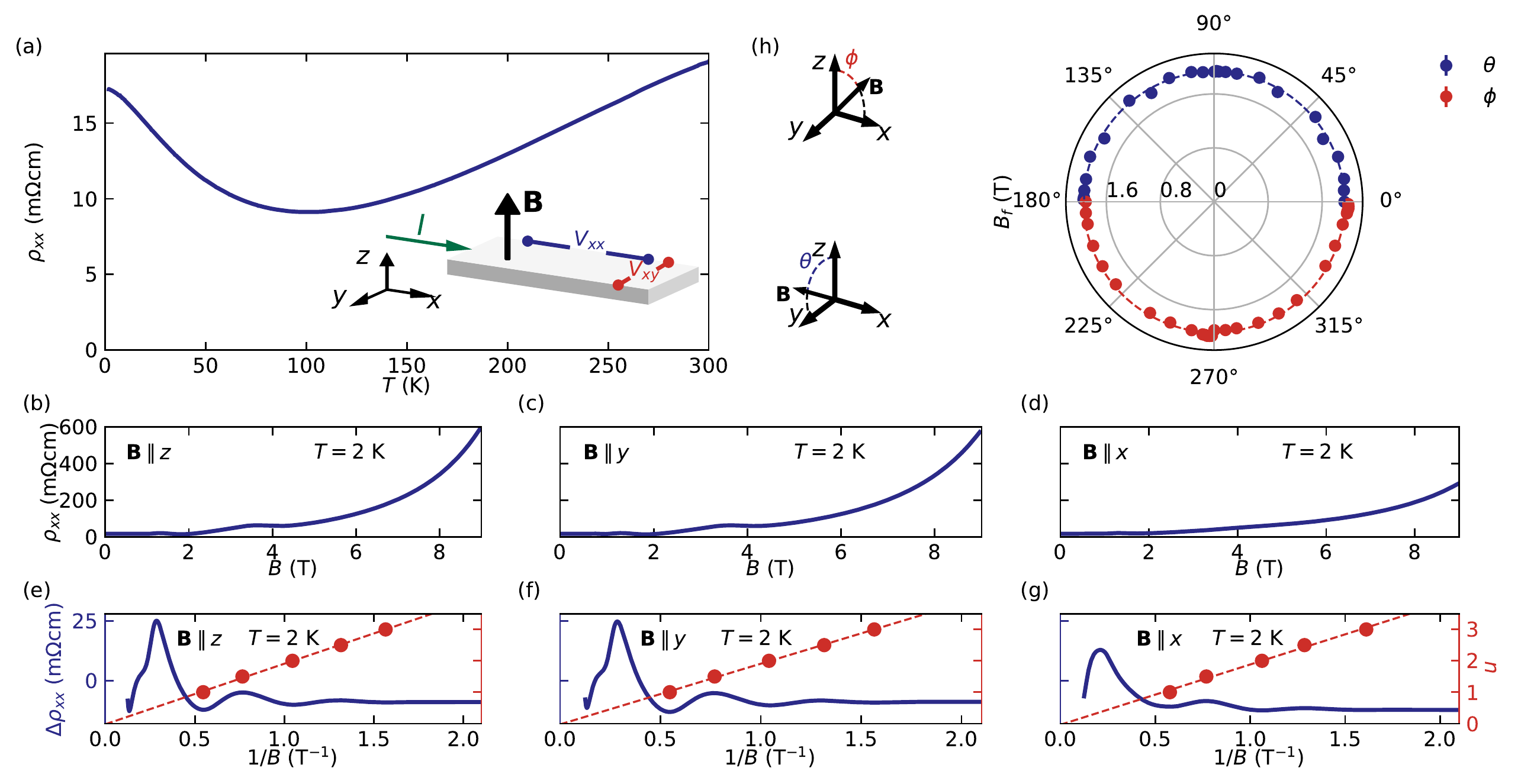}
\caption{\label{fig:fig1}(a) Temperature dependence of longitudinal resitivity~($\rho_{xx}$) of InAs. The inset shows the measurement configuration. (b-d) Field dependence of $\rho_{xx}$ for the three main configurations. (e-g) Oscillatory part of $\rho_{xx}(1/B)$, retrieved by subtraction of a power law function. The red dots mark the positions of the SdH minima and maxima with a dashed line illustrating a linear fit of the Landau indices positions. (h) Frequency of the SdH oscillations as a function of the angle between magnetic field and the direction normal to the plane of the sample. The insets show the two axes around which the rotations were performed. The data was symmetrized accordingly to the cubic symmetry of the crystal structure.}
\end{figure}

InAs takes form of single crystals exhibiting the cubic zinc-blende structure~\cite{adachi1999}. For our electrical transport experiments, mm-size rectangular samples were cleaved out of a $0.5$~mm-thick InAs wafer. The long edges of the samples, defining $x$ and $y$, are parallel to the two $[110]$-type directions of the crystal lattice and their height $z$ is aligned with the $[001]$. We measured two components of the resistivity tensor, namely longitudinal $\rho_{xx}\propto{}V_{xx}/I$ and Hall resistivity $\rho_{xy}\propto{}V_{xy}/I$ (the inset of Fig.~\ref{fig:fig1a}) of 2 samples (referred later as \texttt{A} an \texttt{B}) as a function of temperature $T$ and magnetic field $B$. 

At $T=2$~K, both samples exhibit $\rho_{xx}\sim{}17$~m$\Omega$cm, with a nominal electron density $n=1.6\times{}10^{16}$~cm$^{-3}$ and Hall mobility $\mu{}=2.4\times{}10^{4}$~cm$^{2}$V$^{-1}$s$^{-1}$. All values above are consistent with slight electron $n$-type character. Upon cooling in zero magnetic field, our samples display a metallic-like drop in $\rho_{xx}(T)$ at high temperatures, followed by a rise possibly due to freeze-out of thermally activated shallow donor states, below $100$~K (Fig.~\ref{fig:fig1a}). At lowest temperatures $\rho_{xx}(T)$ saturates resembling impurity scattering limited regime in metals. Almost no change is observed down to 70 mK~\cite{si}. All investigated samples show similar electrical transport properties. In the main text, we focus on data obtained from sample \texttt{A}. Additional data of sample \texttt{B} can be found in the Supplementary Material~\cite{si}. 

To characterize the Fermi surface morphology of our InAs samples and exclude the influence of 2D surface states~\cite{yuan2019} on our transport experiments, we measured Shubnikov-de Haas~(SdH) oscillations with the magnetic field applied along different crystallographic directions. The shape of the surface was reconstructed using the Onsager relation, which connects the period of the oscillations with the maximal orbit of the Fermi surface cross-section perpendicular to the magnetic field direction~\cite{ashcroft76}. Specifically, we rotated $B$ in the $z-x$ and $z-y$ planes of our samples, while measuring $\rho_{xx}(B)$ at 2 K at a series of angles (Fig.~\ref{fig:fig1}(b-d)). The oscillatory part of $\rho_{xx}$ (Fig.~\ref{fig:fig1}(e-g)) was obtained by subtraction of a smooth background in form of a power-law function. The cross-section area of the Fermi surface was then determined from linear fits to the positions of the minima and maxima of the SdH oscillations presented as a function of the inverse of magnetic field $1/B$. We find that all extracted band structure parameters are independent of the field direction, which is in agreement with a single 3D spherical Fermi pocket (Fig.~\ref{fig:fig1h}). 

The Shubnikov-de Haas frequency found in sample \texttt{A} is $B_{\textrm{F}}=1.94(\pm{}0.02)$~T for each field direction, resulting in fully isotropic $k_{\textrm{F}}=7.68(\pm{0.04})\times{}10^{-3}$~\AA$^{-1}$, where the errors denote the standard deviation of the $B_{\textrm{F}}$ for different field directions~(Fig.~\ref{fig:fig1h}). The preceding analysis indicates that for our InAs samples the occupancy of only the lowest Landau level~(LLL) is achieved already for the field below $B_{\textrm{LLL}}=4$~T, where the LLL's energy matches the $E_{\textrm{F}}$, regardless of the field direction. As it was shown by calculations~\cite{pidgeon1967}, despite $g\sim{}9$, in case of InAs with such a small charge carrier concentration (Fermi energy close to the conduction band edge), one should not observe the effects of spin splitting on the scheme of LLs, as the Zeeman term for fields corresponding to the $E_{\textrm{F}}$ is negligibly small in comparison with LL splitting. The only spin related feature observed in the SdH oscillations is the strong additional peak in $\rho_{xx}$ at $B\sim{}4$~T vanishing for currents parallel to the field direction. The appearance of this peak is the result of crossing of the Fermi level by one LL of the spin split pair: $\nu{}=0^{+}$~\cite{pavlov1965,efros1965} ($\nu$ marks the LL index) and has been established that its disappearance only for $B\parallel{}x$ is a result of the suppression of spin-flip assisted scattering between the remaining two spin-polarized Landau bands~\cite{pavlov1965,efros1965}. For more detailed discussion of $\rho_{xx}$ at $\nu{}<1$ see the Ref.~\cite{si}.

\begin{figure}[ht]
\subfloat{\label{fig:fig2a}}
\subfloat{\label{fig:fig2b}}
\subfloat{\label{fig:fig2c}}
\includegraphics[width=\columnwidth]{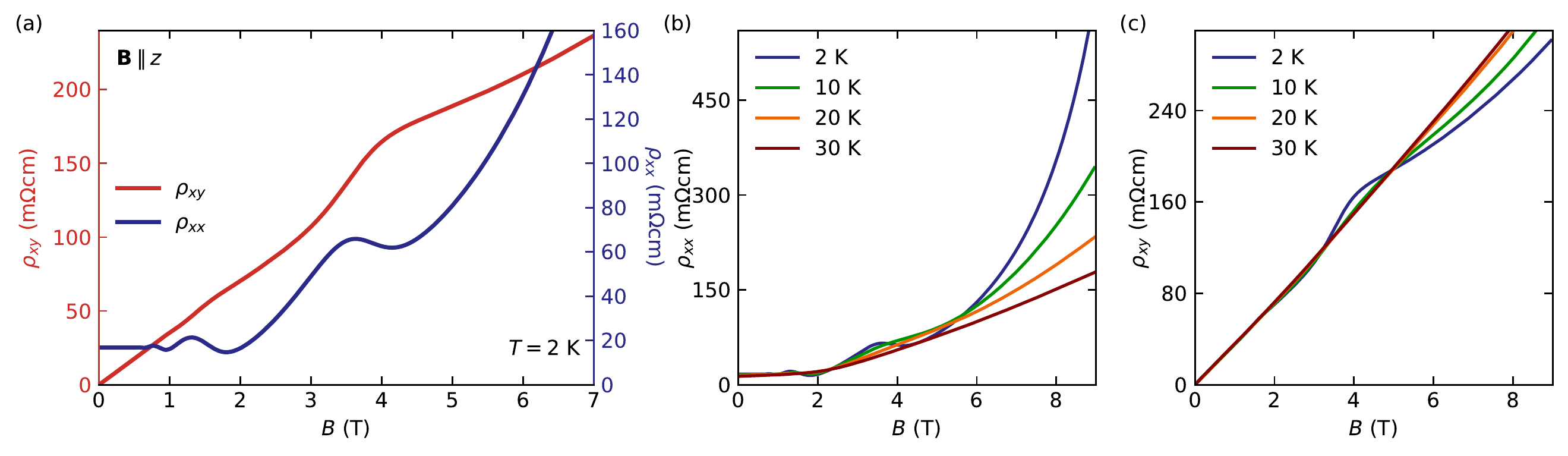}
\caption{\label{fig:fig2}(a) Field dependence of $\rho_{xy}$ and $\rho_{xx}$. (b) and (c) Temperature dependence of $\rho_{xx}(B)$ and $\rho_{xy}(B)$, respectively, measured with $\mathbf{B}\parallel{}\mathbf{z}$.}
\end{figure}

Having confirmed the isotropic 3D electronic Fermi surface in our InAs samples, we next turn to investigate the Hall effect in the configuration with $B$ applied in $z$-direction, that is aligned with the $[001]$ crystallographic axes of the crystal. As shown in Fig.~\ref{fig:fig2a}, we observe signs of plateau-like features in $\rho_{xy}$ that coincide in $B$ with the minima of the SdH oscillations in $\rho_{xx}$ for all field directions, an observation commonly related to the QHE. Both the SdH oscillations in $\rho_{xx}$ and the features in $\rho_{xy}$ are most pronounced at low temperatures, but still visible up to $T=15$~K (Fig.~\ref{fig:fig2}(b-c)).

\section*{Discussion}

\begin{figure}[htb]
\subfloat{\label{fig:fig3a}}
\subfloat{\label{fig:fig3b}}
\begin{center}
\includegraphics[width=0.4\columnwidth]{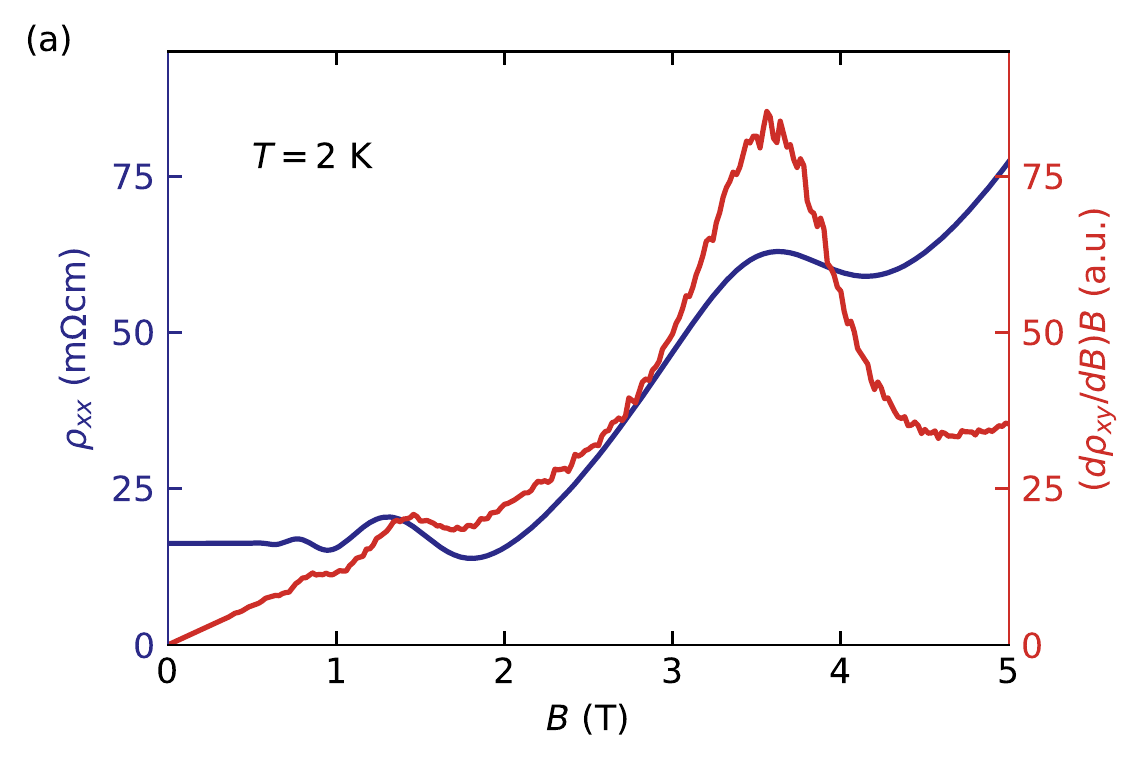}\includegraphics[width=0.4\columnwidth]{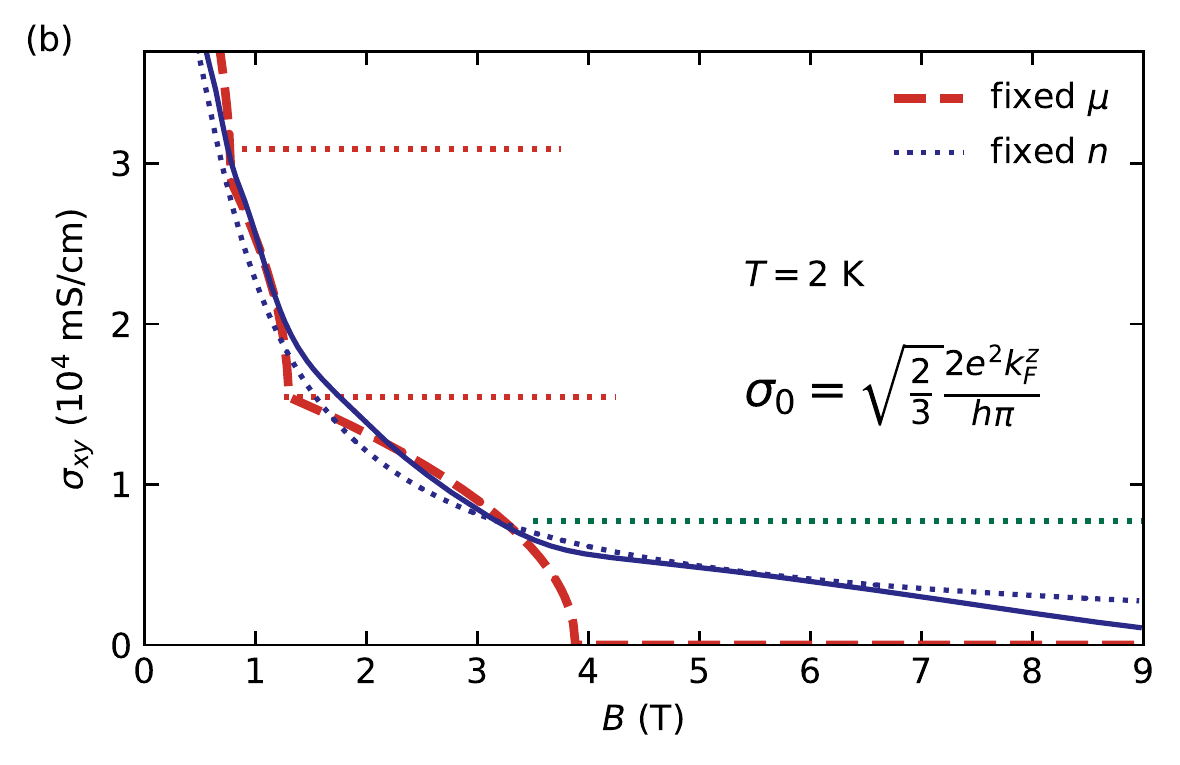}
\end{center}
\caption{\label{fig:fig3}(a) The derivative relation. Field dependence of $\rho_{xx}$ is given in absolute units and field dependence of $(d\rho_{xy}/dB)B$ is scaled to fit the $\rho_{xx}$. (b) Field dependence of $\sigma_{xy}$, the red, dashed line is the result of fit to the Eq.~\ref{eq3}, which implied the fixed chemical potential of charge carriers. The red, dotted lines mark the values of quasi-quantized Hall conductance for $\nu=1$ and $\nu=2$ LL occupancy. The green dotted line marks the contribution from $\nu=0^{-}$ last spin-split Landau level. The dotted blue line shows the result of evaluation performed in conserved particle number regime. Both calculation routines are described in the main text.}
\end{figure}

Qualitative insights into the possible origin of the features observed in $\rho_{xy}$ can be obtained from comparing the shapes of $d\rho_{xy}(B)/dB$ and $\rho_{xx}$. In canonical 2D QHE systems, an empirical observation is that these are connected via $\rho_{xx}(B)=\gamma{}B\cdot d\rho_{xy}(B)/dB$~\cite{chang1985,liu2012}, where $\gamma$ is a dimensionless parameter of the order of $0.01-0.05$, which relates to the local electron concentration fluctuations~\cite{vagner1988,simon1994}. Similar relation was found in both pentatellurides~\cite{galeski2020,galeski20202}. As shown in Fig.~\ref{fig:fig3a}, $d\rho_{xy}(B)/dB$ measured on our InAs samples show maxima and minima at the same field positions as $\rho_{xx}$. In particular, the derivative relation shows strong resemblance with $\gamma=0.35$~(Fig.~\ref{fig:fig3a}), which deviates from the values observed in 2D systems. The factor $\gamma$ taking a very different value in bulk InAs than previously observed in 2D systems cannot be explained due to the lack of fully understanding the nature of the derivative relation itself. In addition, no attempts on generalizing its meaning for 3D systems were undertaken so far. 

A quantitative analysis of the Hall effect in 3D systems based on the resistivity data is not straight forward though. As $\rho_{xx}$ is always finite in our experiments, the Hall conductivity $\sigma_{xy} \neq 1/\rho_{xy}$ and, therefore, we have calculated the Hall conductivity tensor element $\sigma_{xy}=\rho_{xy}/(\rho_{xx}^{2}+\rho_{xy}^{2})$, shown in Fig.~\ref{fig:fig3b}. The example of ZrTe$_{5}$ encouraged analyzing the scenario, where the electrons' chemical potential does not vary with the magnetic field strength. Among the two counterexamples of systems, namely one with fixed charge carrier concentration and one with fixed chemical potential of those charge carriers, the necessity of charge neutrality would require the first case scenario. However, as it is shown in~Fig.~\ref{fig:fig3b}, this would result in completely featureless $\sigma_{xy}=en/B$ (or $\sigma_{xy}=n\mu^{2}B$ for $\mu{}B\ll{}1$), what is in contrast with experimental data.

The analytical calculation for materials with a 3D parabolic band and fixed chemical potential results in~\cite{noky2020}

\begin{equation}\label{eq3}
\sigma_{xy}=\frac{2e^{2}}{h}\frac{1}{2\pi}\sum_{\nu{}=0}^{\infty}2\mathcal{C}_{\nu}\sqrt{k_{\textrm{F}}^{2}-\left(\nu+\frac{1}{2}\right)\frac{2eB}{\hbar}},
\end{equation}

 where $\mathcal{C}_{\nu}=1$ is the Chern number of the parabolic LL bands\cite{ippoliti18,price20}. Despite the fact that, $k_{\textrm{F}}$ is the only system specific parameter in Eq.~\ref{eq3}, which is retrieved here from the frequency of the SdH oscillations, the results of the calculations are in good quantitative agreement with the experimental Hall conductivity slightly beyond the field $B_{\textrm{LLL}}$ up to the point where only the lowest spin-split Landau level ($\nu{}=0^{-}$) is occupied. Further investigations reveals that the feature $\sigma_{xy}$ at $\nu{}=1$ scales with $\sigma_{0}=2e^2/h\cdot{}\sqrt{2/3}k_{F}$ at the onset of the next Landau level, as expected by Ref~\cite{noky2020}. 

As the scenario of $\mu$ moving freely and charge carrier concentration being constant, gives estimates close to both calculations with Eq.~\ref{eq3} and the measured conductivity it completely misses the details of $\sigma_{xy}(B)$. These are captured by theoretical model of fixed $\mu$ employed in this work. However, the model breaks down at the fields where the bottom of the last populated LL is approaching the chemical potential ($B_{\textrm{LLL}}$) what should completely suppress $\sigma_{xy}$, due to Fermi level entering the gap. In reality the prediction and the observation bifurcate at the point where the conductivity value is close to contribution expected for last spin-split LL with index $\nu{}=0^{-}$~(Fig.~\ref{fig:fig3b}). Above this field the fixed-$n$ scenario provides a good agreement, but in a very limited range. This suggest, that the investigated system seems not follow perfectly neither of the two cases (fixed $E_{textrm{F}}$ or fixed $n$) for all fields, but balances in between them. .

Our analysis shows that the model of the quasi-quantized Hall conductivity for isotropic 3D electron systems with a parabolic band proposed in Ref.~\cite{noky2020} is in good agreement with the experimental data close to the LLL of  bulk $n$-type InAs. The mechanism proposed in Ref.~\cite{noky2020} neither depends on the particular purity level of the sample nor its shape and is rather an intrinsic property of the 3D electronic structure. Compared to earlier attempts to explain the generically observed features close to the LLL of doped III-V semiconductors~\cite{mani1990,viehweger1991,zeitler1994}, this makes the model of the quasi-quantized Hall conductivity a more comprehensive explanation of the experiments. However, we would like to emphasize that these different explanations do not necessarily contradict each other but some are rather complementary.

\section*{Conclusion}

In summary, we have show that the features observed in the Hall measurements of InAs are in qualitative and quantitative agreement with the existence of the quasi-quantized Hall signal in an isotropic 3D electron system. Our findings render the Hall effect in InAs is another flavor of the effect observed in Dirac semimetals~\cite{galeski20202}. It is likewise rooted in the mixture of common features of the band structure and the field-dependence of the chemical potential. The requirements to observe the quasi-quantized Hall effect in 3D materials are low charge carrier density. Hence, we propose that doped semiconductors and semimetals are ideal future candidates for its observation. This example shows the necessity of careful treatment of observed anomalies in Hall conductance, especially in systems with low charge carrier concentrations. Of special interest also would be a generalization of the 3D quasi-quantized Hall response to materials with higher Chern numbers, or to the anomalous-type Hall effects in magnetic Weyl semimetals.

\section*{Methods}

Electrical transport measurements employed low-frequency~($f=77.777$~Hz) lock-in technique. An excitation current of $I_{\textrm{e}}=100$~$\mu$A~(peak-to-peak value) was applied, giving the current density of the order of $j\sim{}3$~mA/cm$^{2}$.

\bibliography{bib}

\section*{Acknowledgements}

Authors would like to thank Elena Hassinger for fruitful discussions.

\section*{Author contributions statement}

J.G. conceived the experiment,  R.W. and S.G. conducted the magnetoresistance measurements, R.W. and J.G. analyzed the data, J.N., Y.S. and J.G provided the model for calculations. R.W., J.G. and C.F. wrote the manuscript. All authors reviewed the manuscript. 

\section*{Competing interests}

The authors declare no competing interests.

\newpage

\section*{Additional information}

\renewcommand{\thefigure}{S\arabic{figure}}
\setcounter{figure}{0}

\subsection*{Hall conductivity and derivative relation for Sample B.}

\begin{figure}[ht]
\begin{center}
\includegraphics[width=0.4\columnwidth]{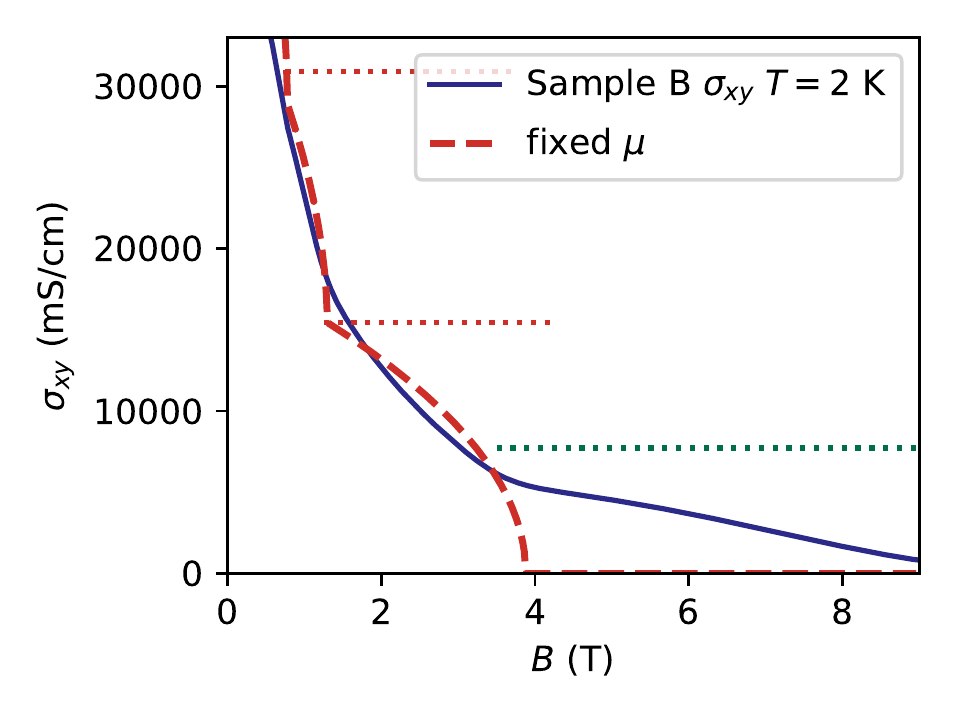}
\end{center}
\caption{\label{fig:figs1}Field dependence of $\sigma_{xy}$ for sample \texttt{B}. The red, dashed line is the result of fit to the Eq.~\ref{eq3}, which implied the fixed chemical potential of charge carriers. The red, dotted lines mark the values of quasi-quantized Hall conductance for $\nu=1$ and $\nu=2$ filling factors. The green dotted line marks the contribution from $\nu=0^{-}$ last spin-split Landau level.}
\end{figure}

\begin{figure}[ht]
\begin{center}
\includegraphics[width=0.4\columnwidth]{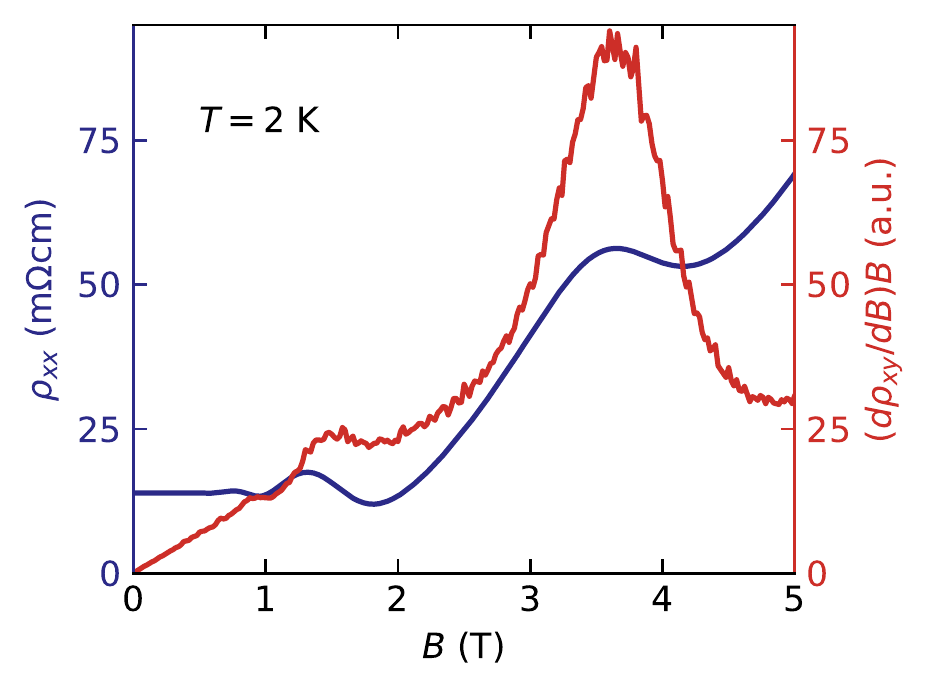}
\end{center}
\caption{\label{fig:figs1a}The derivative relation for Sample B. Field dependence of $\rho_{xx}$ is given in absolute units and field dependence of $(d\rho_{xy}/dB)B$ is scaled to fit the $\rho_{xx}$. I n case of Sample B parameter $\gamma=0.36$.}
\end{figure}

\subsection*{Longitudinal magnetoresistance beyond $\mathbf{B}_{\textrm{LLL}}$.}

\begin{figure}[ht]
\begin{center}
\includegraphics[width=0.4\columnwidth]{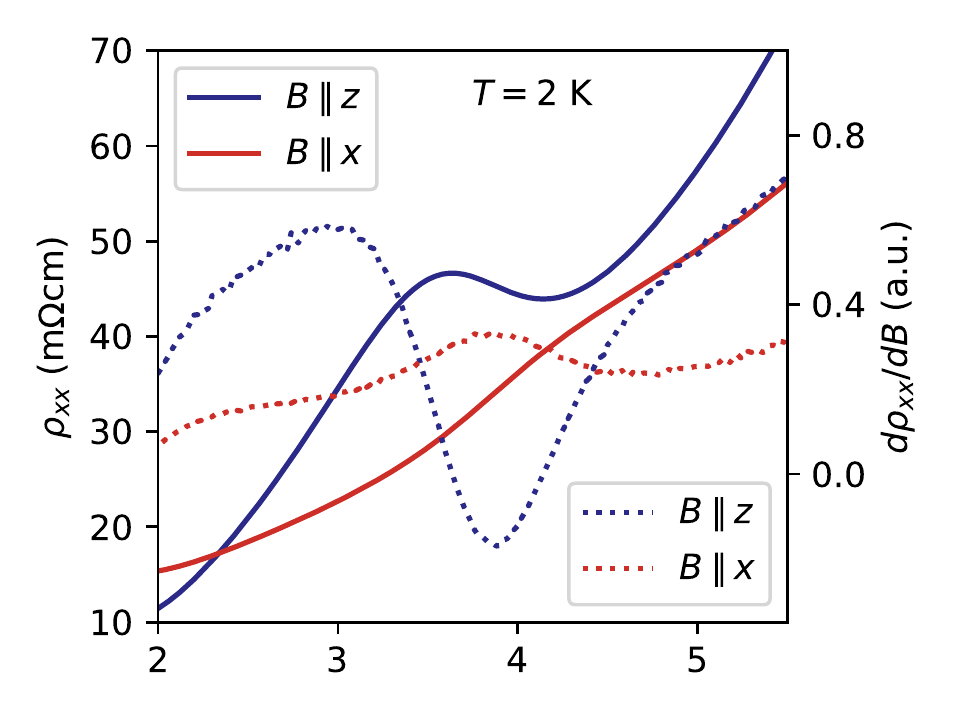}
\end{center}
\caption{\label{fig:figs2}Field dependence of $\rho_{xx}$ and its derivative, around $B_{\textrm{LLL}}$, for field directions perpendicular and parallel to the electric current.}
\end{figure}

In Figs.~\ref{fig:figs1}(e-g) showing the oscillatory part of $\rho_{xx}(B)$, apart from the maximum marking the $\mu=0^{+}$ spin-split LL one can determine the presence of much smaller shoulder placed in slightly higher fields. Careful examination of $\rho_{xx}$ curves and their derivatives, in the vicinity of $B_{\textrm{LLL}}$~(Fig.~\ref{fig:figs2}), shows that for field directions not parallel to the direction of electric current flow no signs of such a shoulder is distinguishable in the $\rho_{xx}$. This would suggest the bump at the side of $\nu=0^{+}$ peak being an artifact resulting from power-law background subtraction, which does not capture increase in rate change of resistivity in field after crossing $B_{\textrm{LLL}}$, what could be expected. 

On the other hand, $\rho_{xx}$ for $B\parallel{}x$ does show a bump, easily noticeable thanks to the suppression of $\nu=0^{+}$ peak in that configuration. This peak appears in the fields outside of range covered by the theory employed in this work and at this point we cannot account for this feature. We hope that future measurements at sub-Kelvin temperatures, with use of dilution refrigerator, could allow us, by suppressing thermal broadening of energy levels to resolve observed feature in greater detail and address this open question.

\subsection*{QQHE in Sample A at low temperatures.}

In addition we measure the resitivity of our Sample A at low temperatures using dilution refrigerator for sample \texttt{A}. No significant difference was observed. The only additional feature appearing in Fig.~\ref{fig:figs5} is placed at the value of conductivity corresponding to contribution from three spin-split LLs ($\nu=0^{-}$, $\nu=0^{+}$ and $\nu=1^{-}$). Lower temperatures should be more favorable towards its observation, as thermal smearing of LLs is smaller. However it is hard to confirm its origin with confidence.


\begin{figure}[ht]
\begin{center}
\includegraphics[width=0.45\columnwidth]{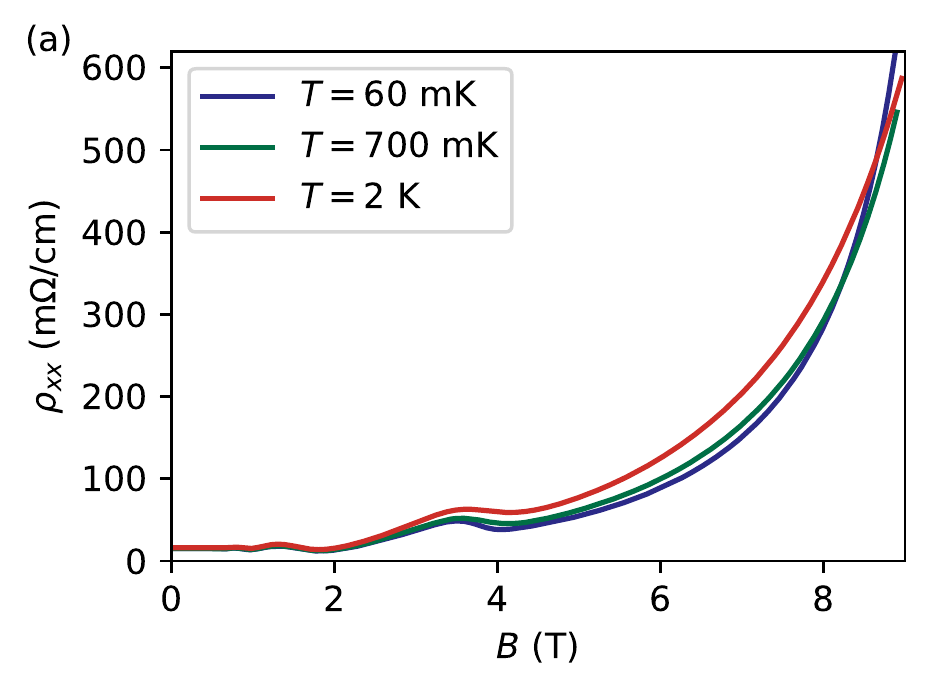}\includegraphics[width=0.45\columnwidth]{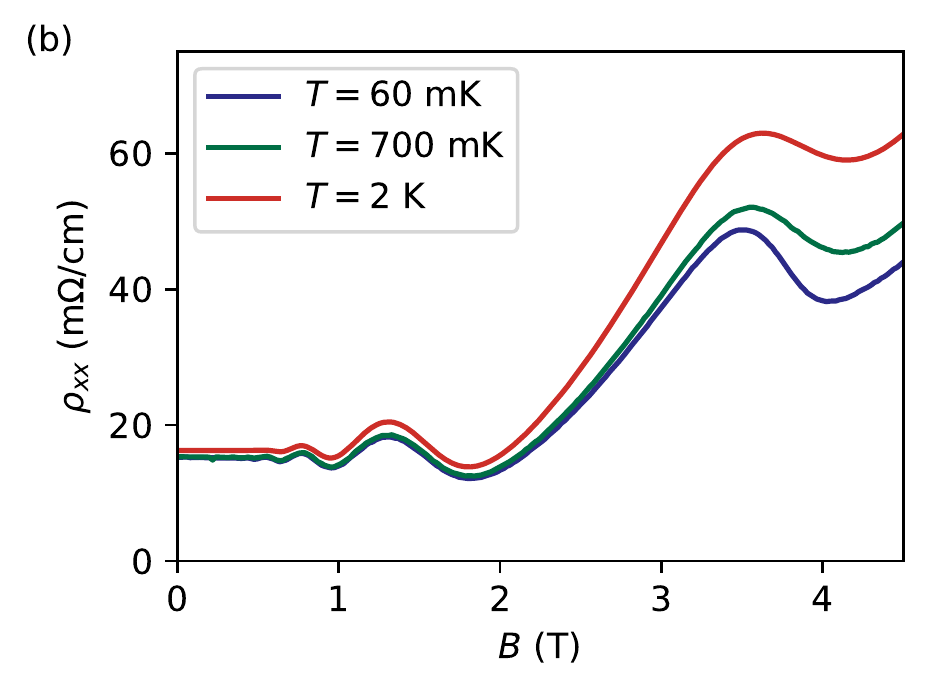}
\end{center}
\caption{\label{fig:figs3}(a) Field dependence of $\rho_{xx}$ at different temperatures. (b) Low-field part of the data, where SdH oscillations can be observed.}
\end{figure}

\begin{figure}[ht]
\begin{center}
\includegraphics[width=0.4\columnwidth]{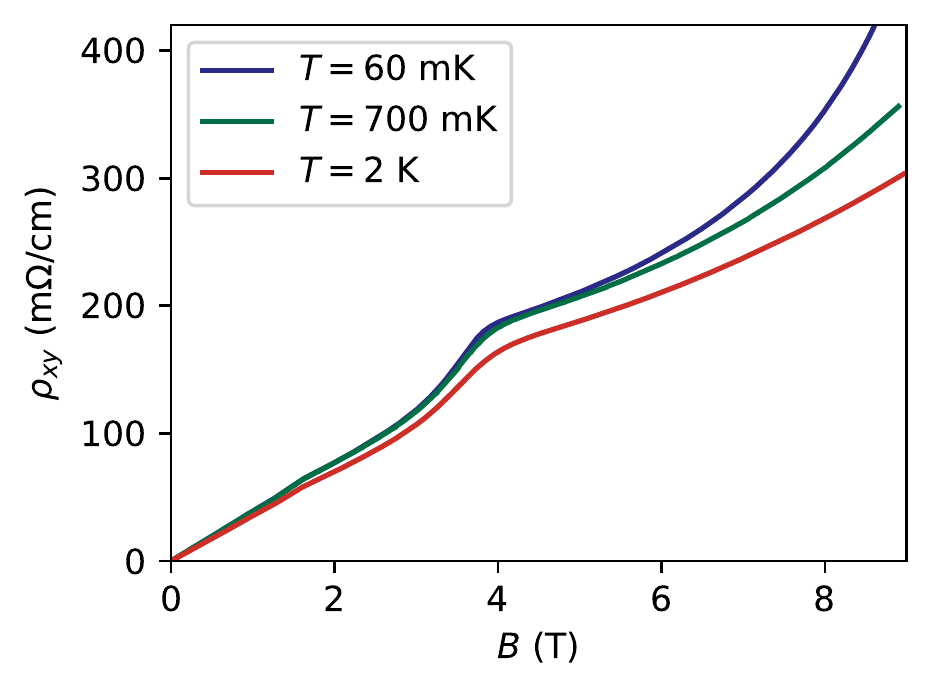}
\end{center}
\caption{\label{fig:figs4}Field dependence of $\rho_{xy}$ at different temperatures.}
\end{figure}

\begin{figure}[ht]
\begin{center}
\includegraphics[width=0.4\columnwidth]{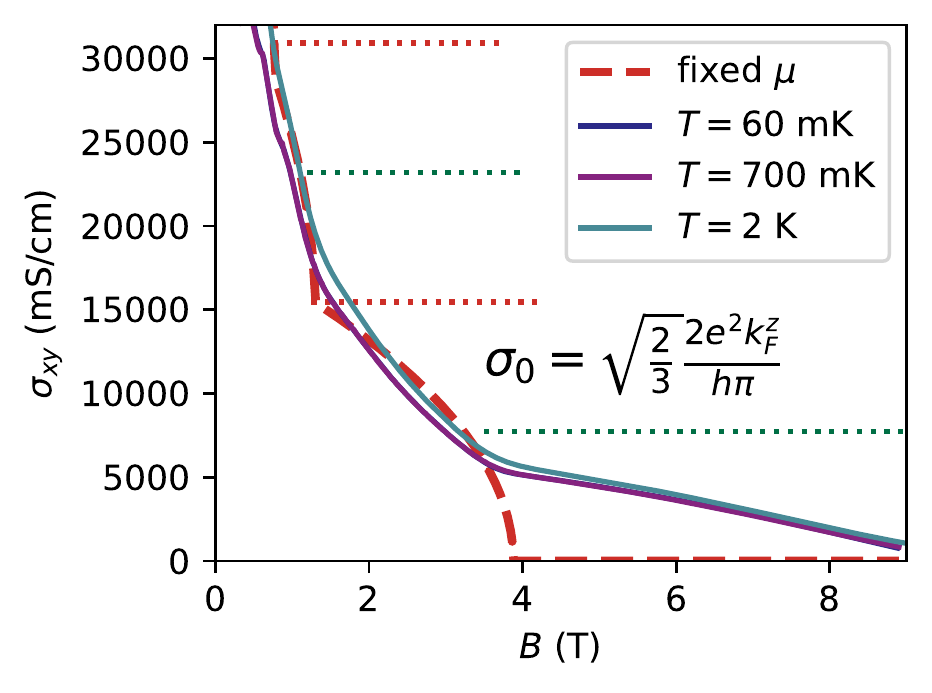}
\end{center}
\caption{\label{fig:figs5}Field dependence of $\sigma_{xy}$, the red, dashed line is the result of fit to the Eq.~\ref{eq3} from the main text, which implied the fixed chemical potential of charge carriers. The red, dotted lines mark the values of quasi-quantized Hall conductance for $\nu=1$ and $\nu=2$ filling factors. The green dotted lines mark the contribution from $\nu=0^{-}$ last spin-split Landau level and three lowest spin split levels, possible contribution of which might be faintly visible in the lowest temperatures.}
\end{figure}

\end{document}